\documentclass[prb,preprint]{revtex4-1}%
\usepackage{amsfonts}
\usepackage{amsmath}
\usepackage{amssymb}
\usepackage{graphicx}%
\newtheorem{theorem}{Theorem}
\newtheorem{acknowledgement}[theorem]{Acknowledgement}

\DeclareMathOperator{\Hermite}{H}
\DeclareMathOperator{\Laguerre}{L}
\DeclareMathOperator{\g}{g}
\DeclareMathOperator{\G}{G}
\DeclareMathOperator{\Ei}{Ei}
\DeclareMathOperator{\Tr}{Tr}
\newcommand{\half}{\mathchoice{{\textstyle{\frac12}}}{{\textstyle{\frac12}}}{{\scriptstyle{1/2}}}{{\scriptscriptstyle{1/2}}}}
\begin{document}
\preprint{ }
\title{Time-dependent quantum transport in a resonant tunnel junction coupled to a
nanomechanical oscillator}
\author{M. Tahir}
\author{A. MacKinnon}
\email[E-Mail: ]{a.mackinnon@imperial.ac.uk}
\affiliation{The Blackett Laboratory, Imperial College London, South
Kensington campus, London SW7 2AZ, U.K.}

\begin{abstract}
We present a theoretical study of time-dependent quantum transport in a
resonant tunnel junction coupled to a nanomechanical oscillator within the
non-equilibrium Green's function technique. An arbitrary voltage is applied to
the tunnel junction and electrons in the leads are considered to be at zero
temperature. The transient and the steady state behavior of the system is
considered here in order to explore the quantum dynamics of the oscillator as
a function of time. The properties of the phonon distribution of the
nanomechnical oscillator strongly coupled to the electrons on the dot are
investigated using a non-perturbative approach. We consider both the energy
transferred from the electrons to the oscillator and the Fano factor as a
function of time. We discuss the quantum dynamics of the nanomechanical
oscillator in terms of pure and mixed states. We have found a significant
difference between a quantum and a classical oscillator. In particular, the
energy of a classical oscillator will always be dissipated by the electrons
whereas the quantum oscillator remains in an excited state. This will provide
useful insight for the design of experiments aimed at studying the quantum
behavior of an oscillator.

\end{abstract}

\maketitle

\section{Introduction}
Nanoscopic physics has been a subject of increasing experimental and
theoretical interest for its potential applications in nanoelectromechanical
systems (NEMS)\cite{1,2,3}. The physical properties of these devices are of
crucial importance in improving our understanding of the fundamental science
in this area including many-body phenomena\cite{4}. One of the most striking
paradigms exhibiting many body effects in mesoscopic science is quantum
transport through single electronic levels in quantum dots and single
molecules\cite{5,6,7,8} coupled to external leads. Realizations of these
systems have been obtained using semiconductor beams coupled to single
electron transistors (SET's) and superconducting single electron transistors
(SSET's)\cite{9,10}, carbon nanotubes\cite{11} and, most recently, suspended
graphene sheets\cite{12}. Such systems can be used as a direct measure of
small displacements, forces and mass in the quantum regime. The quantum
transport properties of these systems require extremely sensitive measurement
that can be achieved by using SET's, or a resonant tunnel junction, and
SSET's. In this context, NEMS are not only interesting devices studied for
ultrasensitive transducers but also because they are expected to exhibit
several exclusive features of transport phenomena such as avalanche-like
transport and shuttling instability\cite{13,14}. The nanomechanical properties
of a resonant tunnel junction coupled to an oscillator\cite{15} or a
SET\cite{16,17} coupled to an oscillator are currently playing a vital role in
enhancing the understanding of NEMS.

The nanomechanical oscillator coupled to a resonant tunnel junction or SET is
a close analogue of a molecule being used as a sensor whose sensitivity has
reached the quantum limit\cite{1,2,3,9,18}. The signature of quantum states
has been predicted for the nanomechanical oscillator coupled to the
SET's\cite{9} and SSET's\cite{10,19}. In these experiments, it has been
confirmed that the nanomechanical oscillator is strongly affected by the
electron transport in the circumstances where we are also trying to explore
the quantum regime of NEMS. In this system, electrons tunnel from one of the
leads to the isolated conductor and then to the other lead. Phonon assisted
tunneling of non--resonant systems has mostly been shown by experiments on
inelastic tunneling spectroscopy (ITS). With the advancement of modern
technology, as compared to ITS, scanning tunneling spectroscopy (STS) and
scanning tunneling microscopy (STM) have proved more valuable tools for the
investigation and characterization of molecular systems\cite{20} in the
conduction regime. In STS\ experiments, significant signatures of the strong
electron-phonon interaction have been observed\cite{21,22} beyond the
established perturbation theory. Hence, a theory beyond master equation
approach or linear response is necessary. Most of the theoretical work on
transport in NEMS has been done within the scattering theory approach
(Landauer) but it disregards the contacts and their effects on the scattering
channel as well as effect of electrons and phonons on each other\cite{23}.
Very recently, the non--equilibrium Green's function (NEGF)
approach\cite{24,25,26} has been growing in importance in the quantum
transport of nanomechanical systems\cite{15,16,17,18,27,28}. An advantage of
this method is that it treats the infinitely extended reservoirs in an exact
way\cite{29}, which may lead to a better understanding of the essential
features of NEMS. NEGF has been applied in the study of shot noise in chain
models\cite{30} and disordered junctions\cite{31} while noise in Coulomb
blockade Josephson junctions has been discussed within a phase correlation
theory approach\cite{32}. In the case of an inelastic resonant tunneling
structure, in which strong electron-phonon coupling is often considered, a
very strong source-drain voltage is expected for which coherent electron
transport in molecular devices has been considered by some workers\cite{33}
within the scattering theory approach. Inelastic effects on the transport
properties have been studied in connection with NEMS and substantial work on
this issue has been done, again within the scattering theory approach\cite{23}%
. Recently, phonon assisted resonant tunneling conductance has been discussed
within the NEGF technique at zero temperature\cite{34}. To the best of our
knowledge, in all these studies, time-dependent quantum transport properties
of a resonant tunnel junction coupled to a nanomechanical oscillator have not
been discussed so far. The development of time-dependent quantum transport for
the treatment of nonequilibrium system with phononic as well as Fermionic
degree of freedom has remained a challenge since the 1980's\cite{35}.
Generally, time-dependent transport properties of mesoscopic systems without
nanomechanical oscillator have been reported\cite{36} and, in particular,
sudden joining of the leads with quantum dot molecule have been
investigated\cite{35,37} for the case of a noninteracting quantum dot and for
a weakly Coulomb interacting molecular system. Strongly interacting systems in
the Kondo regime have been investigated\cite{38,39}. More recently\cite{40},
the transient effects occurring in a molecular quantum dot described by an
Anderson-Holstein Hamiltonian has been discussed. To this end, we present the
following study.

In the present work, we shall investigate the time evolution of a quantum dot
coupled to a single vibrational mode as a reaction to a sudden joining to the
leads. We employ the non-equilibrium Green's function method in order to
discuss the transient and steady state dynamics of NEMS. This is a fully
quantum mechanical formulation whose basic approximations are very
transparent, as the technique has already been used to study transport
properties in a wide range of systems. In our calculation inclusion of the
oscillator is not perturbative as the STS experiments\cite{21,22} are beyond
the perturbation theory. So a non-perturbative approach is required beyond the
quantum master equation\cite{27,28,41} or linear response. Hence, our work
provides an exact analytical solution to the current--voltage characteristics,
including coupling of leads with the system, very small chemical potential
difference and both the right and left Fermi level response regimes. For
simplicity, we use the wide-band approximation\cite{25,35,42,43}, where the
density of states in the leads and hence the coupling between the leads and
the dot is taken to be independent of energy. Although the method we are using
does not rely on this approximation. This provides a way to perform transient
transport calculations from first principles while retaining the essential
physics of the electronic structure of the dot and the leads. Another
advantage of this method is that it treats the infinitely extended reservoirs
in an exact way in the present system, which may give a better understanding
of the essential features of NEMS in a more appropriate quantum mechanical picture.

\section{Model calculations}
We consider a single quantum dot connected to two identical metallic leads. A
single oscillator is coupled to the electrons on the dot and the applied gate
voltage is used to tune the single level of the dot. In the present system, we
neglect the spin degree of freedom and electron-electron interaction effects
and consider the simplest possible model system. We also neglect the effects
of finite electron temperature of the lead reservoirs and damping of the
oscillator. Our model consists of the individual entities such as the single
quantum dot and the left and right leads in their ground states at zero
temperature. The Hamiltonian of our simple system\cite{34,42,43} is%
\begin{equation}
H_{\mbox{\scriptsize dot-ph}}=\left[  \epsilon_{0}+\lambda l(b^{\dagger
}+b)\right]  c_{0}^{\dag}c_{0}+\hbar\omega(b^{\dagger}b+{\half})\,,
\label{1}%
\end{equation}
where $\epsilon_{0}$ is the single energy level of electrons on the dot with
$c_{0}^{\dag},c_{0}$ the corresponding creation and annihilation operators,
the coupling strength, $\eta=\lambda l,$ with $\lambda=eE,$ is the
electrostatic field between electrons on the dot and an oscillator, seen by
the electrons due to the charge on the oscillator, $l=\sqrt{\hbar/
2m\omega}$ is the zero point amplitude of the oscillator, $\omega$ is the
frequency of the oscillator and $b^{\dagger},b$ are the raising and lowering
operator of the phonons. The remaining elements of the Hamiltonian are%

\begin{eqnarray}
H_{\mbox{\scriptsize leads}}&=&\sum_{j}\epsilon_{j}c_{j}^{\dagger}c_{j},
\label{2}\\%
H_{\mbox{\scriptsize leads-dot}}&=&\frac{1}{\sqrt{N}}\sum_{j}V_{\alpha
}(t)\left(  c_{j}^{\dagger}c_{0}+c_{0}^{\dagger}c_{j}\right)  , \label{3}%
\end{eqnarray}
where we include time-dependent hopping $V_{\alpha}(t)$\ to enable us to
connect the leads $\alpha=L,R$\ to the dot at a finite time. For the
time-dependent dynamics, we shall focus on sudden joining of the leads to the
dot at $t=0$, which means $V_{\alpha}(t)=V\theta(t)$, where $\theta(t)$ is the
Heaviside unit step function. $N$ is the total number of states in the lead,
and $j$ represents the channels in one of the leads. For the second lead the
Hamiltonian can be written in the same way. The total Hamiltonian of the
system is thus $H=H_{\mbox{\scriptsize dot-ph}}+H_{\mbox{\scriptsize leads}}%
+H_{\mbox{\scriptsize leads-dot}}\,$. We write the eigenfunctions of
$H_{\mbox{\scriptsize dot-ph}}$ as%
\begin{eqnarray}
\Psi_{m}(K,x_{0}\neq0)&=&A_{m}\exp[-{\textstyle\frac{l^{2}K^{2}}{2}}%
]\Hermite_{m}(lK)\exp[-\mathrm{i} Kx_{0}] \label{4}\\%
\Psi_{n}(K,x_{0}=0)&=&A_{n}\exp[-{\textstyle\frac{l^{2}K^{2}}{2}}]\Hermite_{n}(lK)\,,
\label{5}%
\end{eqnarray}
for the occupied, $x_{0}\not =0$ and unoccupied, $x_{0}=0$, dot respectively,
where $x_{0}={\lambda}/{2m\omega^{2}}$ is the shift of the oscillator due
to the coupling to the electrons on the dot, where $A_{n}={1}/{\sqrt
{\sqrt{\pi}2^{n}n!l}}$, $A_{m}={1}/{\sqrt{\sqrt{\pi}2^{m}m!l}}$, and
$\Hermite_{n}(lK)$ are the usual Hermite polynomials. Here we have used the fact that
the harmonic oscillator eigenfunctions have the same form in both real and
Fourier space ($K$).

In order to transform between the representations for the occupied and
unoccupied dot we require the matrix with elements $\Phi_{nm}=\int\Psi
_{n}^{\ast}(K,x_{0}=0)\Psi_{m}(K,x_{0}\neq0)\,{\mathrm{d}}K,$ which may be
simplified\cite{44} as%
\begin{eqnarray}
\Phi_{n,m}  &=& \frac{l}{\sqrt{\pi2^{m+n}n!m!}}\int\exp\left(  -l^{2}%
K^{2}\right) \Hermite_{n}^{\ast}(lK)\Hermite_{m}(lK)\exp\left(\mathrm{i}Kx_{0}\right)
\,{\mathrm{d}}K\label{6}\\
&=& \sqrt{\frac{2^{m-n}n!}{m!}}\exp\left(  -{\textstyle\frac{1}{4}}%
x^{2}\right)  \left(  {\half}\mathrm{i}x\right)  ^{m-n}%
\Laguerre_{n}^{m-n}\left(  {\half}{x^{2}}\right)  \label{7}%
\end{eqnarray}
for $n\leq m$, where $x={x_{0}}/{l}$ and $\Laguerre_{n}^{m-n}(x)$ are the
associated Laguerre polynomials. Note that the integrand is symmetric in $m$
and $n$ but the integral is only valid for $n\leq m$. Clearly the result for
$n>m$ is obtained by exchanging $m$ and $n$ in equation~(\ref{7}) to obtain
\begin{equation}
\Phi_{n,m}=\sqrt{\frac{2^{|m-n|}\min[n,m]!}{\max[n,m]!}}\exp\left(
-\textstyle\frac{1}{4}x^{2}\right)  \left(  \half%
\mathrm{i}x\right)  ^{|m-n|}\Laguerre_{\min[n,m]}^{|m-n|}\left(  \textstyle\frac{1}%
{2}{x^{2}}\right)  \,. \label{8}%
\end{equation}

In order to calculate the analytical solutions and to discuss the numerical
results of the transient and steady state dynamics of the nanomechanical
systems, our focus in this section is to derive an analytical relation for the
time dependent effective self-energy and the Green's functions. In obtaining
these results we use the wide--band approximation only for simplicity,
although the method we are using does not rely on this approximation, where
the retarded self--energy of the dot due to each lead is given by\cite{25,35}
\begin{equation}
\Sigma_{\alpha}^{r}(t_{1},t_{2})=V_{\alpha}^{\ast}(t_{1})\g_{\alpha,\alpha}%
^{r}(t_{1},t_{2})V_{\alpha}(t_{2}), \label{9}%
\end{equation}
where $\alpha=L,R$ represent the left and right leads and the Green's function
in the leads for the uncoupled system is%

\[
\g_{\alpha,\alpha}^{r}(t_{1},t_{2})=\frac{1}{N}\sum_{j}\g_{\alpha,j}^{r}%
(t_{1},t_{2})=-\mathrm{i}n_{\alpha}\theta(t_{1}-t_{2})\overset{+\infty}{\underset
{-\infty}{%
%TCIMACRO{\dint }%
%BeginExpansion
{\displaystyle\int}
%EndExpansion
}}\mathrm{d}
\varepsilon_{\alpha}\exp[-\mathrm{i}\varepsilon_{\alpha}(t_{1}-t_{2})],
\]
with the fact that $\underset{j}{%
%TCIMACRO{\dsum }%
%BeginExpansion
{\displaystyle\sum}
%EndExpansion
}\mapsto\overset{+\infty}{\underset{-\infty}{%
%TCIMACRO{\dint }%
%BeginExpansion
{\displaystyle\int}
%EndExpansion
}}Nn_{\alpha}\mathrm{d}\varepsilon_{\alpha},$\ where $j$ stands for every channel in
each lead and $n_{\alpha}$ is the constant number density of the leads.

Now using the uncoupled Green's function into equation~(\ref{9}), the retarded self
energy may be written as%

\begin{eqnarray}
\Sigma_{\alpha}^{r}(t_{1},t_{2})  &=& -\mathrm{i}n_{\alpha}\theta(t_{1}-t_{2}%
)\overset{+\infty}{\underset{-\infty}{%
%TCIMACRO{\dint }%
%BeginExpansion
{\displaystyle\int}
%EndExpansion
}}\mathrm{d}\varepsilon_{\alpha}V_{\alpha}^{\ast}(t_{1})\exp[-\mathrm{i}\varepsilon_{\alpha
}(t_{1}-t_{2})]V_{\alpha}(t_{2}),\label{10}\\
&=& -\mathrm{i}n_{\alpha}V_{\alpha}^{\ast}(t_{1})V_{\alpha}(t_{2})\theta(t_{1}%
-t_{2})\overset{+\infty}{\underset{-\infty}{%
%TCIMACRO{\dint }%
%BeginExpansion
{\displaystyle\int}
%EndExpansion
}}\mathrm{d}\varepsilon_{\alpha}\exp[-\mathrm{i}\varepsilon_{\alpha}(t_{1}-t_{2})],\nonumber\\
&=& -\mathrm{i}n_{\alpha}V_{\alpha}^{\ast}(t_{1})V_{\alpha}(t_{2})\theta(t_{1}%
-t_{2})\times2\pi\delta(t_{1}-t_{2}), \label{11}%
\end{eqnarray}
Now we use the fact that $V_{\alpha}(t_{1})=\left\vert V\right\vert
\times\theta(t_{1})$, $V_{\alpha}(t_{2})=\left\vert V\right\vert \times
\theta(t_{2})$. Then the above expression can be written as%

\begin{equation}
\Sigma_{\alpha}^{r}(t_{1},t_{2})=-\half\mathrm{i}\Gamma_{\alpha}\theta
(t_{2})\delta(t_{1}-t_{2}) \label{12}%
\end{equation}
where $\Gamma_{\alpha}=4\pi\left\vert V\right\vert ^{2}n_{\alpha}$ is the
damping factor ($\Gamma_{L}=\Gamma_{R}=\Gamma$). Similarly $\Sigma_{\alpha
}^{a}(t_{1},t_{2})=\left[  \Sigma_{\alpha}^{r}(t_{1},t_{2})\right]  ^{\ast
}=+\half\mathrm{i}\Gamma_{\alpha}$ $\theta(t_{2})\delta(t_{1}-t_{2})\,\,.$

We solve Dyson's equation using $H_{\text{dot-leads}}$, as a
perturbation. In the presence of the oscillator, the retarded and advanced
Green's functions on the dot, with the phonon states in the representation of
the unoccupied dot, may be written as
\begin{equation}
\G_{n,n^{\prime}}^{r}(t,t_{1})=\sum_{m}\Phi_{n,m}\g_{m}^{r}(t,t_{1}%
)\Phi_{n^{\prime},m}^{\ast}\,,\text{ }\G_{n,n^{\prime}}^{a}(t_{2},t^{\prime
})=\sum_{k}\Phi_{n,k}\g_{k}^{a}(t_{2},t^{\prime})\Phi_{n^{\prime},k}^{\ast}
\label{13}%
\end{equation}
where $\g_{m(k)}^{r(a)}$ is the retarded (advanced) Green's function on the
occupied dot coupled to the leads may be written as,%
\begin{eqnarray}
\g_{m}^{r}(t,t_{1}) &=& -\mathrm{i}\theta(t-t_{1})\times\exp[-\mathrm{i}(\varepsilon_{m}%
-\mathrm{i}\Gamma)(t-t_{1})],\text{ }t_{1}>0 \label{14}\\
\g_{k}^{a}(t_{2},t^{\prime}) &=&+\mathrm{i}\theta(t_{2}-t^{\prime})\times\exp[-\mathrm{i}(\varepsilon
_{k}+\mathrm{i}\Gamma)(t_{2}-t^{\prime})],\text{ \ \ }t_{2}>0 \label{15}%
\end{eqnarray}
with $\varepsilon_{m}=\epsilon_{0}+(m+\half)\hbar\omega-\Delta$,
$\varepsilon_{k}=\epsilon_{0}+(k+\half)\hbar\omega-\Delta$, and $\Delta
=\lambda^{2}/2m{\omega}^{2}$.

The above Eqs.~(\ref{12}, \ref{13}, \ref{14}, \ref{15}) will be the starting point of our
examination of the time-dependent response of the coupled system. These
functions are the essential ingredients for theoretical considerations of such
diverse problems as low and high voltage, coupling of electron and phonons,
transient and steady state phenomena.

\section{Time-dependent dot population $\rho(t)$}
The density matrix is related to the dot population through $\rho
(t)=\underset{n}{\sum}\rho_{n,n}(t,t)$, where the density matrix $\rho
_{n,n}(t,t)=-\mathrm{i}\G_{n,n^{\prime}}^{<}(t,t^{\prime})$, for $t=t^{\prime}$ and
$n=n^{\prime}$. $\G_{n,n^{\prime}}^{<}(t,t^{\prime})$ is the lesser
Green's function\cite{24,25,35} on the dot including all the contributions
from the leads. The lesser Green's function for the dot in the presence of the
nanomechanical oscillator is given by%
\begin{equation}
\G_{n,n^{\prime}}^{<}(t,t^{\prime})=\sum_{n_{0},n_{0}^{\prime},\alpha}%
%TCIMACRO{\dint }%
%BeginExpansion
{\displaystyle\int}
%EndExpansion%
%TCIMACRO{\dint }%
%BeginExpansion
{\displaystyle\int}
%EndExpansion
\mathrm{d}t_{1}\mathrm{d}t_{2}\G_{n,n_{0}}^{r}(t,t_{1})\Sigma_{n_{0},n_{0}^{\prime},\alpha}%
^{<}(t_{1},t_{2})\G_{n_{0}^{\prime},n^{\prime}}^{a}(t_{2,}t^{\prime}),\text{
\ \ }t\text{ and }t^{\prime}>0 \label{16}%
\end{equation}
whereas, for $t$ and $t^{\prime}<0$, the $\G_{n,n^{\prime}}^{<}(t,t^{\prime})$
is equal to zero, and $G_{n,n^{\prime}}^{<}(t,t^{\prime})$ includes all the
information of the nanomechanical oscillator and electronic leads of the
system, and $n_{0},n_{0}^{\prime},n,n^{\prime}$ are the oscillator indices.
The lesser self-energy, $\Sigma_{n_{0},n_{0}^{\prime},\alpha}^{<}(t_{1}%
,t_{2})$, contains electronic and oscillator contributions. The electronic
contributions are non-zero only when $t_{1}$ and $t_{2}>0$. As the oscillator
is initially in its ground state, only the $n_{0}=n_{0}^{\prime}=0$ term gives
a non-zero contribution to the lesser self-energy. The lesser self--energy for
the dot may be written as%

\[
\Sigma_{0,0,\alpha}^{<}(t_{1},t_{2})=V_{\alpha}^{\ast}(t_{1})\g_{\alpha,\alpha
}^{<}(t_{1},t_{2})V_{\alpha}(t_{2}),
\]
with%

\[
\g_{\alpha,\alpha}^{<}(t_{1},t_{2})=\frac{1}{N}\underset{j}{%
%TCIMACRO{\dsum }%
%BeginExpansion
{\displaystyle\sum}
%EndExpansion
}\g_{\alpha,j}^{<}(t_{1},t_{2})=\overset{+\infty}{\underset{-\infty}{%
%TCIMACRO{\dint }%
%BeginExpansion
{\displaystyle\int}
%EndExpansion
}}\mathrm{d}\varepsilon_{\alpha}{\mathop{\mathrm{f}}\nolimits}_{\alpha}(\varepsilon_{\alpha})2\mathrm{i}n_{\alpha}%
\exp[-\mathrm{i}\varepsilon_{\alpha}(t_{1}-t_{2})],
\]
where ${\mathop\mathrm{f}\nolimits}_{\alpha}(\varepsilon_{\alpha})$ is the Fermi distribution functions
of the left and right leads, which have different chemical potentials under a
voltage bias. For the present case of zero temperature the lesser self--energy
may be recast in terms of the Heaviside step function $\theta(x)$ as
\begin{equation}
\Sigma_{0,0,\alpha}^{<}(t_{1},t_{2})=\mathrm{i}\Gamma_{\alpha}\overset
{+\infty}{\underset{-\infty}{%
%TCIMACRO{\dint }%
%BeginExpansion
{\displaystyle\int}
%EndExpansion
}}\frac{\mathrm{d}\varepsilon_{\alpha}}{2\pi}\theta\left(  \epsilon_{\mathrm{F\alpha}%
}+\half\hbar\omega-\varepsilon_{\alpha}\right)  \theta(t_{1}%
)\theta(t_{2})\exp[-\mathrm{i}\varepsilon_{\alpha}(t_{1}-t_{2})]\,, \label{17}%
\end{equation}
where $\Sigma_{0,0,\alpha}^{r,(a),(<)}(t_{1},t_{2})$ are all non-zero only
when both times ($t_{1},t_{2}$) are positive $t_{1},t_{2}>0$ and
$\epsilon_{\mathrm{F\alpha}}$ is the Fermi energy on each of leads.

The density matrix $\rho_{n,n}(t,t)$ can be calculated by using Eqs.~(\ref{12},
\ref{13}, \ref{14}, \ref{15}, \ref{17}) in Eq.~(\ref{16}) at $t=t^{\prime}$ and $n=n^{\prime}$ as%

\begin{eqnarray*}
\rho_{n,n}(t,t)  &=& -\mathrm{i}\sum_{\alpha,m,k}%
%TCIMACRO{\dint _{0}^{t}}%
%BeginExpansion
{\displaystyle\int_{0}^{t}}
%EndExpansion%
%TCIMACRO{\dint _{0}^{t}}%
%BeginExpansion
{\displaystyle\int_{0}^{t}}
%EndExpansion
\mathrm{d}t_{1}\mathrm{d}t_{2}\Phi_{n,m}\Phi_{0,m}^{\ast}\exp[-\mathrm{i}(\varepsilon_{m}-\mathrm{i}\Gamma
)(t-t_{1})]\\
&&  \times\{\mathrm{i}\Gamma%
%TCIMACRO{\dint _{-\infty}^{\epsilon_{\mathrm{F}\alpha}}}%
%BeginExpansion
{\displaystyle\int_{-\infty}^{\epsilon_{\mathrm{F}\alpha}}}
%EndExpansion
\frac{\mathrm{d}\varepsilon_{\alpha}}{2\pi}\exp[-\mathrm{i}\varepsilon_{\alpha}(t_{1}%
-t_{2})\,\}\Phi_{0,k}\Phi_{n,k}^{\ast}\exp[-\mathrm{i}(\varepsilon_{k}+\mathrm{i}\Gamma
)(t_{2}-t)],
\end{eqnarray*}
Although $\g^{r,(a)}(t_{1},t_{2})$ is non-zero for $t<0$, it is never required
due to the way it combines with $\Sigma_{0,0,\alpha}^{r,(a),(<)}(t_{1},t_{2}%
)$. By carrying out the time integrations, the resulting expression is written as%
\begin{eqnarray*}
\rho_{n,n}(t,t)  &=& \frac{\Gamma}{2\pi}\sum_{\alpha,m,k}%
%TCIMACRO{\dint _{-\infty}^{\epsilon_{\mathrm{F}\alpha}}}%
%BeginExpansion
{\displaystyle\int_{-\infty}^{\epsilon_{\mathrm{F}\alpha}}}
%EndExpansion
\mathrm{d}\varepsilon_{\alpha}\frac{\Phi_{n,m}\Phi_{0,m}^{\ast}\Phi_{0,k}\Phi
_{n,k}^{\ast}}{(\varepsilon_{\alpha}-\varepsilon_{k}-\mathrm{i}\Gamma)(\varepsilon
_{\alpha}-\varepsilon_{m}+\mathrm{i}\Gamma)}\\
&&  \times\left\{1+\exp[\mathrm{i}(\varepsilon_{k}-\varepsilon_{m}+2\mathrm{i}\Gamma
)t]-\exp[-\mathrm{i}(\varepsilon_{\alpha}-\varepsilon_{k}-\mathrm{i}\Gamma)t]-\exp
[\mathrm{i}(\varepsilon_{\alpha}-\varepsilon_{m}+\mathrm{i}\Gamma)t]\right\}
\end{eqnarray*}
The integral over the energy in the above equation is carried out\cite{45}.
The final result for the density matrix is written as%
\begin{equation}
\rho_{n,n}(t,t)=\frac{\Gamma}{2\pi}\sum_{m,k}\frac{\Phi_{n,m}\Phi_{m,0}^{\ast
}\Phi_{0,k}\Phi_{n,k}^{\ast}}{\varepsilon_{k}-\varepsilon_{m}+2\mathrm{i}\Gamma}%
[Y^{L}_{mk}+Y^{R}_{mk}+Z^{L}_{mk}+Z^{R}_{mk}], \label{18}%
\end{equation}
where we have added the contribution from the right and the left leads, which
can be written in terms of$\ \alpha$ as%

\begin{eqnarray*}
Y^{\alpha}_{mk}  &=&\left\{1+\exp[\mathrm{i}(\varepsilon_{k}-\varepsilon_{m}%
+2\mathrm{i}\Gamma)t]\right\}
\left\{
\ln(\epsilon_{\mathrm{F}\alpha}-\varepsilon_{k}-\mathrm{i}\Gamma) 
-\ln(\epsilon_{\mathrm{F}\alpha}-\varepsilon_{m}+\mathrm{i}\Gamma)
\right\}\\
 &=&\left\{1+\exp[\mathrm{i}(\varepsilon_{k}-\varepsilon_{m}%
+2\mathrm{i}\Gamma)t]\right\}\\
&& \times\left\{\half\frac{\ln[(\epsilon_{\mathrm{F}\alpha
}-\varepsilon_{k})^{2}+\Gamma^{2}]}{\ln[(\epsilon_{\mathrm{F}\alpha
}-\varepsilon_{m})^{2}+\Gamma^{2}]}%
+\mathrm{i}\left[\tan^{-1}\left(\frac{\varepsilon_{F\alpha}-\varepsilon_{k}}{\Gamma}%
\right)+\tan^{-1}\left(\frac{\varepsilon_{F\alpha}-\varepsilon_{m}}{\Gamma}\right)+\pi\right]\right\},\\
Z^{\alpha}_{mk}  &=& \exp[\mathrm{i}(\varepsilon_{k}-\varepsilon_{m}+2\mathrm{i}\Gamma
)t]\left\{-\Ei[\mathrm{i}(\epsilon_{\mathrm{F}\alpha}-\varepsilon_{k}%
-\mathrm{i}\Gamma)t]+\Ei[-\mathrm{i}(\epsilon_{\mathrm{F}\alpha}-\varepsilon
_{m}+\mathrm{i}\Gamma)t]\right\}\\
&&  +\left\{\Ei[\mathrm{i}(\epsilon_{\mathrm{F}\alpha}-\varepsilon_{m}%
+\mathrm{i}\Gamma)t]-\Ei[-\mathrm{i}(\epsilon_{\mathrm{F}\alpha}-\varepsilon
_{k}-\mathrm{i}\Gamma)t]\right\},
\end{eqnarray*}
with $\epsilon_{\mathrm{F}\alpha}$\ being the right and the left Fermi levels
and $\Ei(x)$ the exponential integral function. Special care is required in
evaluating the $\Ei(x)$ to choose the correct Riemann sheets in order to make
sure that these functions are consistent with the initial conditions
$\rho(0)=0$\ and are continuous functions of time and chemical potential.  The same applies to complex logarithms in the first, apparently simpler, form for $Y^{\alpha}_{mk}$.

Now using equation (\ref{18}), the dot population may be written as%
\[
\rho(t)=\underset{n}{\sum}\rho_{n,n}(t,t)=\frac{\Gamma}{2\pi}\sum_{n,m,k}%
\frac{\Phi_{n,m}\Phi_{m,0}^{\ast}\Phi_{0,k}\Phi_{n,k}^{\ast}}{\varepsilon
_{k}-\varepsilon_{m}+2i\Gamma}[Y^{L}_{mk}+Y^{R}_{mk}+Z^{L}_{mk}+Z^{R}_{mk}].
\]

\section{Time-dependent Current from lead $\alpha$}
The particle current $I_{\alpha}$ into the interacting region from the lead is
related to the expectation value of the time derivative of the number operator
$N_{\alpha}=\sum_{\alpha j}c_{\alpha j}^{\dagger}c_{\alpha j},$
as\cite{25,35,36,37}%

\begin{equation}
I_{\alpha}=-e\left<\frac{\mathrm{d}}{\mathrm{d}t}x\right>
=\frac{-\mathrm{i}e}{\hbar}\left<[H,x]\right> \label{19}%
\end{equation}
and the final result for the current through each of the leads is written as
(See appendix~\ref{App.A})

\begin{equation}
I_{\alpha}(t)=\frac{e\Gamma}{2\pi\hbar}\sum_{m}\Phi_{0,m}\Phi_{0,m}^{\ast
}\left\{I^{1\alpha}_m+I^{2L}_m+I^{2R}_m\right\}, \label{20}%
\end{equation}
where
\begin{eqnarray*}
I^{1\alpha}_m &=& 2\left(\tan^{-1}\left[\frac{\epsilon_{\mathrm{F}\alpha}-\varepsilon_{m}%
}{\Gamma}\right]+\frac{\pi}{2}\right)\\
&& -\mathrm{i}\left\{
 \Ei[+\mathrm{i}(\epsilon_{\mathrm{F}\alpha}-\varepsilon_{m}+\mathrm{i}\Gamma)t]
-\Ei[-\mathrm{i}(\epsilon_{\mathrm{F}\alpha}-\varepsilon_{m}-\mathrm{i}\Gamma)t]\right\}  ,\\
I^{2\alpha}_m  &=& -\left(  1+\exp[-2\Gamma t]\right)  \left(\tan^{-1}%
\left[\frac{\varepsilon_{F\alpha}-\varepsilon_{m}}{\Gamma}\right]+\frac{\pi}{2}\right)\\
&&  -\half\mathrm{i}\exp[-2\Gamma t]\left\{
 \Ei[+\mathrm{i}(\epsilon_{\mathrm{F}\alpha}-\varepsilon_{m}-\mathrm{i}\Gamma)t]
-\Ei[-\mathrm{i}(\epsilon_{\mathrm{F}\alpha}-\varepsilon_{m}+\mathrm{i}\Gamma)t]
\right\}\\
&&  +\half\mathrm{i}\left\{
 \Ei[+\mathrm{i}(\epsilon_{\mathrm{F}\alpha}-\varepsilon_{m}+\mathrm{i}\Gamma)t]
-\Ei[-\mathrm{i}(\epsilon_{\mathrm{F}\alpha}-\varepsilon_{m}-\mathrm{i}\Gamma)t]
\right\},
\end{eqnarray*}
where in calculating the left current we need $I^{1L}_m$ and both the
contributions $I^{2L}_m$ and $I^{2R}_m$ and for the right current $I^{1L}_m$ is replaced by $I^{1R}_m$.  As before, special care is required in
evaluating the $\Ei(x)$ to choose the correct Riemann sheets in order to make
sure that these functions are consistent with the initial conditions
$I_{\alpha}(t)=0$\ and are continuous functions of time and chemical potential.

\section{Average energy and Fano factor}
To calculate the energy transferred from the electrons to the nanomechanical
oscillator, we return to the density matrix $\rho_{n,n}(t,t)$ given in Eq.
(\ref{18}). We may therefore use the lesser Green's function or density matrix to
calculate the energy transferred to the oscillator as%
\begin{equation}
E_{ph}=\langle n\hbar\omega\rangle=\frac{\underset{n}{\sum}n\hbar\omega\rho_{n,n}%
(t,t)}{\underset{n}{\sum}\rho_{n,n}(t,t)}. \label{21}%
\end{equation}
Note that the normalisation in equation~(\ref{21}) is required as the bare density matrix contians both electronic and oscillator contributions.  The trace eliminates the oscillator part, leving the electronic part.
In order to further characterize the state of the nanomechanical oscillator we
investigate the Fano factor for the change of average occupation number,
$<n>$ as a function of time. The corresponding relation for the Fano factor
is given by\cite{46}%
\begin{equation}
F=\frac{\langle n^{2}\rangle-\langle n\rangle^{2}}{\langle n\rangle},
\label{22}%
\end{equation}
where $\langle n\rangle=
\underset{n}{\sum}n\rho_{n,n}(t,t)/{\underset{n}{\sum}\rho_{n,n}(t,t)}$ and $\langle n^{2}\rangle=
{\underset{n}{\sum}n^{2}\rho_{n,n}(t,t)}/{\underset{n}{\sum}\rho_{n,n}(t,t)}$,   with the average evaluated using
the diagonal element of the density matrix on the quantum dot.

\section{Discussion of Results}
The dot population, net current through the system, total current into the
system, average energy and Fano factor of a resonant tunnel junction coupled
to a nanomechanical oscillator are shown graphically as a function of time for
different values of coupling strength, tunneling rate, and voltage bias. The
following parameters\cite{1,2,3,4,5,6,7,8,9,10,13,15,16,17,18,19,32} were
employed: the single energy level of the dot $\epsilon_{0}=0.5$, and the
characteristic frequency of the oscillator $\hbar\omega=0.1$. These
parameters will remain fixed for all further discussions and have same
dimension as of $\hbar\omega$. We are interested in small and large values
of tunneling from the leads, different values of the coupling strength between
the electrons and the nanomechanical oscillator, and of the left chemical
potential $0\leq\epsilon_{\mathrm{F}L}\leq1$. 

\begin{figure}[h!tb]
%\begin{minipage}{\linewidth}
%\begin{center}
\includegraphics[width=0.5\textwidth]{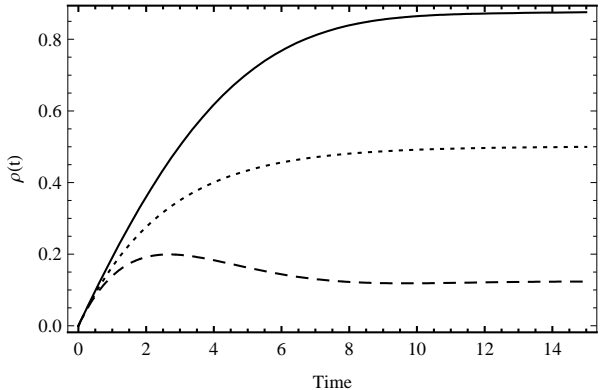}
%\end{center}
\caption{\label{fig.1}Time-dependent dot population $\rho(t)$ against time for different
pairs of the right and the left Fermi energies (0,0), (0,1), (1,1). The dotted
line correspond to empty, dashed line correspond to half full and solid line
corresponds to almost full state of the dot. Parameters: $\epsilon
_{0}=0.5,\hbar\omega=0.1,\Gamma=0.1,\eta=0.05$ . Units: all the parameters
have same dimension as of $\hbar\omega$.}
%\end{minipage}\\
\end{figure}
\begin{figure}[h!tb]
%\begin{minipage}{\linewidth}
%\begin{center}
\includegraphics[width=0.5\textwidth]{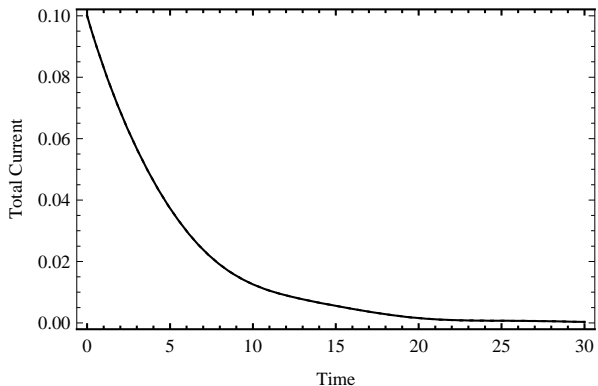}
%\end{center}
\caption{\label{fig.2}Total current ($I_{L}(t)+I_{R}(t)$) flowing onto the dot as a
function of time for fixed values of $\epsilon_{0}=0.5,\hbar\omega
=0.1,\Gamma=0.1,\eta=0.05,\epsilon_{FR}=0,\epsilon_{FL}=1.$This current (solid
line) is equivalent to the rate of change of dot population $\frac{d}{dt}%
\rho(t)$ (dashed line) as a function of time for same parameters as of
current. In this figure, solid and dashed lines have same values at all
points. Units: all the parameters have same dimension as of $\hbar\omega$.}
%\end{minipage}
\end{figure}

The nanomechanical oscillator
induced resonance effects are clearly visible in the numerical results. It
must be noted that we have obtained these results in the regime of \ both
strong and zero or weak coupling between the nanomechanical oscillator and the
electrons on the dot. The tunneling of electrons between the leads and the dot
is considered to be symmetric ($\Gamma_{R}=\Gamma_{L}$) and we assume that the
leads have constant density of states.

The dot population is shown in fig.~\ref{fig.1}, as a function of time in order to see
the transient and steady state dynamics of the system. We consider here empty,
half full and occupied states of the system for fixed values of $\Gamma=0.1,$
$\eta=0.05,$ by choosing the right and the left Fermi levels pairs ( 0, 0),
(0, 1) and (1, 1) respectively. Firstly, when both the Fermi levels are below
the dot energy then the dot population rises initially for a short time and
for long times settles at a small but finite value. This is not quite empty
because the finite $\Gamma$\ allows some tunneling onto the dot. Secondly,
when the left Fermi level is above the dot energy then the dot population
settles in a partially full (half full) state. Thirdly, when both the Fermi
levels are above the dot energy, it is completely full for a short time but
for long time is not quite full, again due to the dot coupling with the leads.
These results are consistent with the particle-hole symmetry of the system as
the empty state of the system is not empty and the occupied state is not
completely full, while the partially full is roughly half full.

In fig.~\ref{fig.2}, we have shown the total current flowing onto the dot as a function
of time for fixed values of $\Gamma=0.1,\eta=0.05,\epsilon_{\mathrm{F}R}=0,$
and of the left Fermi level 1. This current (solid line) is equivalent to the
rate of change of the dot population (dashed line) for the same parameters. In
this figure, we can not distinguish the solid and the dashed line. This
confirms that our analytical results are consistent with the equation of
continuity, $I_{L}(t)+I_{R}(t)=\frac{d}{dt}\rho(t)$, and hence, with the
conservation laws for all parameters.

\begin{figure}[htb]
\begin{center}
\includegraphics[width=0.7\textwidth]{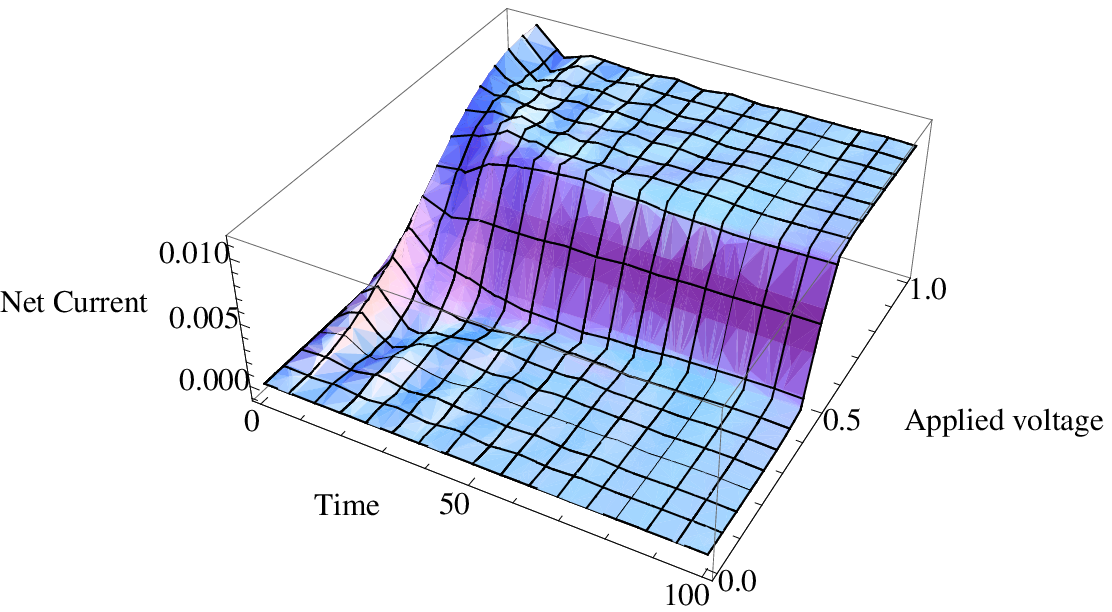}
\includegraphics[width=0.7\textwidth]{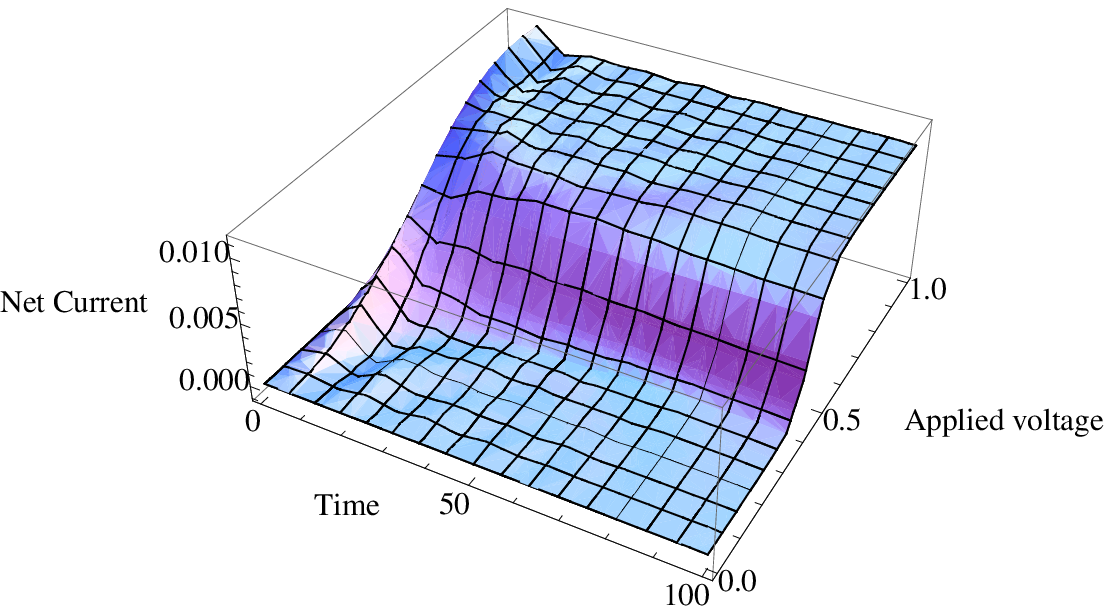}
\end{center}
\caption{\label{fig.3}Net current ($I_{L}(t)-I_{R}(t)$) flowing through the system as a
function of both time and of the left Fermi level for two different values of
coupling strength: $\eta=0.02$ (Fig.~\ref{fig.3}(a))$,$ and $0.1$ (Fig.~\ref{fig.3}(b)). Parameters:
$\epsilon_{0}=0.5,\epsilon_{FR}=0,\epsilon_{FL}=1,\hbar\omega=0.1,\Gamma
=0.01$. Units: all the parameters have same dimension as of $\hbar\omega$.}
\end{figure}

\begin{figure}[h!tb]
\begin{center}
\includegraphics[width=0.5\textwidth]{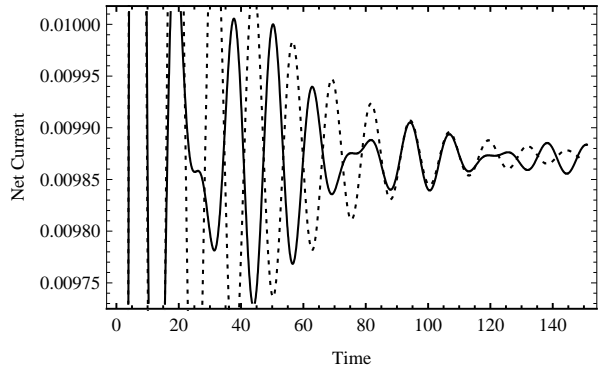}
\end{center}
\caption{\label{fig.4}Net current ($I_{L}(t)-I_{R}(t)$) flowing through the system as a
function of time for two different values of coupling strength: $\eta=0.02$
(dotted line)$,$ and $0.08$ (solid line). Parameters: $\epsilon_{0}%
=0.5,\epsilon_{FR}=0,\epsilon_{FL}=1,\hbar\omega=0.1,\Gamma=0.01$. Units:
all the parameters have same dimension as of $\hbar\omega$.}
\end{figure}

\begin{figure}[h!tb]
\begin{center}
\includegraphics[width=0.5\textwidth]{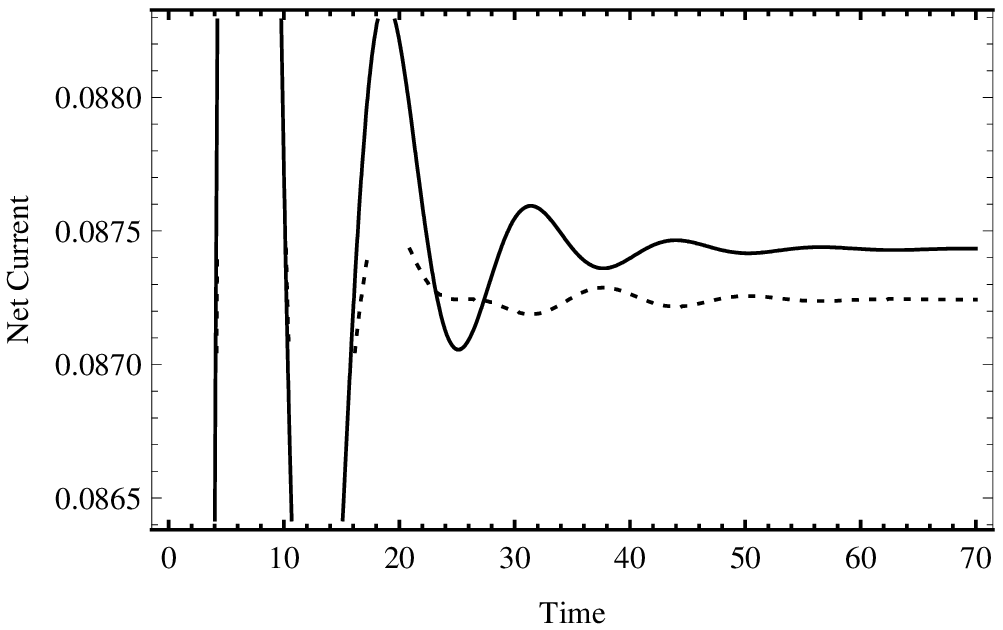}
\end{center}
\caption{\label{fig.5}Net current ($I_{L}(t)-I_{R}(t)$) flowing through the system as a
function of time for two different values of coupling strength: $\eta=0.02$
(dotted line)$,$ and $0.08$ (solid line), and $\Gamma=0.1$. All the parameters
are same as in fig.~\ref{fig.4} and have same dimension as of $\hbar\omega$.}
\end{figure}

In fig.~\ref{fig.3} we have shown the net current ($I_{L}(t)-I_{R}(t)$) flowing through
the system as a function of both time and of the left Fermi level for two
different values of coupling strength: $\eta=0.02$ to $\eta=0.08$ and for
small and large values of $\Gamma$. We observe simple oscillations in the net
current flowing through the system for weak coupling strength and weak
tunneling. With increasing coupling strength the structure of the oscillations
becomes more complicated as shown in fig.~\ref{fig.3}(b). In order to interpret this
complicated structure, we have a two step discussion: firstly, we have plotted
the net current as a function of time in fig.~\ref{fig.4} with fixed values of the Fermi
level, $\epsilon_{\mathrm{F}L}=1$, $\epsilon_{\mathrm{F}R}=0,$ tunneling
energy, $\Gamma=0.01$ and for different values of coupling strength:
$\eta=0.02$ and $\eta=0.08.$ In this figure, in the limit of weak coupling the
oscillations are again simple while for the strong coupling limit, there is a
beating pattern in the oscillations. We note that the frequency of the simple
oscillations is ($\left\vert \epsilon_{\mathrm{F}L}-\epsilon_{0}\right\vert $)
and these oscillations are present even in the limit of weak coupling. We
conclude that this is a purely electronic process (plasmon oscillations). It
is clear from the figure that in the strong coupling case, it contains two
beating frequencies, therefore we interpret this as due to a mixture of
electronic and mechanical frequencies. Secondly, in fig.~\ref{fig.5}, we have plotted the
net current for fixed values of $\epsilon_{\mathrm{F}L}=1$, $\epsilon
_{\mathrm{F}R}=0,$ tunneling energy, $\Gamma=\hbar\omega$ and for different
values of coupling strength: $\eta=0.02$ and $\eta=0.08.$ We have found that
in the regime ( $\Gamma\geq\hbar\omega$), the effects of the oscillator are
not apparent and the period of the nanomechanical oscillator can not be
resolved. Why can the period of the oscillator not be resolved by the
electrons in this limit? In this regime, electrons spend less time on the dot
than the period of the oscillator. Therefore, electrons do not resolve the
period of the nanomechanical oscillator. Now we will focus only in the regime
of small tunneling $\Gamma<\hbar\omega$,\ for further discussion in order to
analyze the dynamics of the nanomechanical oscillator and the effects of
coupling between the electrons and the nanomechanical oscillator.

\begin{figure}[h!tb]
\begin{center}
\includegraphics[width=0.7\textwidth]{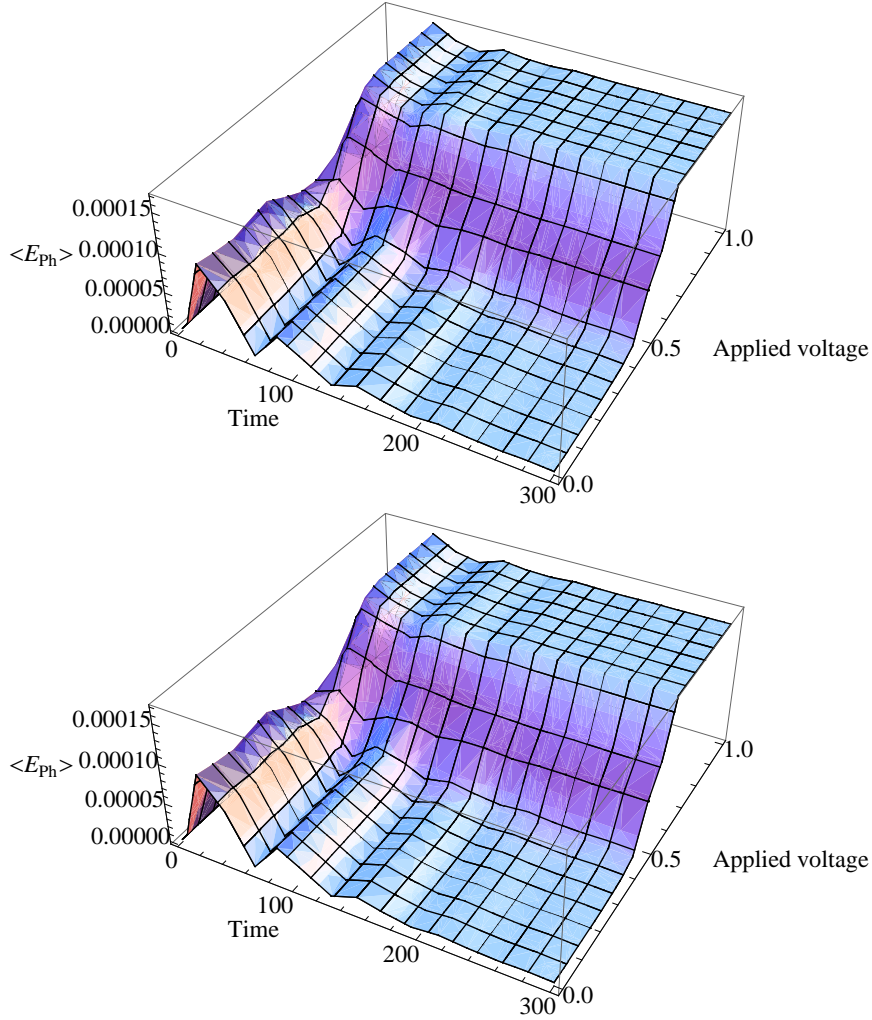}
\includegraphics[width=0.7\textwidth]{Fig.6a.eps}
\end{center}
\caption{\label{fig.6}Average energy transferred to the oscillator as a function of time
and left Fermi level for fixed values of $\epsilon_{0}=0.5,\epsilon
_{FR}=0,\Gamma=0.01$ and for different values of coupling strength:
$\eta=0.02$ (Fig.~\ref{fig.6}(a))$,$ and $0.08$ (Fig.~\ref{fig.6}(b)). Units: all the parameters
have same dimension as of $\hbar\omega$.}
\end{figure}

Next we have shown the average energy of the nanomechanical oscillator as a
function of time and of the left Fermi energy in fig.~\ref{fig.6} for fixed values of
tunneling $\Gamma=0.01$, $\epsilon_{\mathrm{F}R}=0,$ and for different values
of coupling strength $\eta=0.02,$ $\eta=0.08$. We found damped oscillations
for short times and constant energy for long times. This constant average
energy increases with increasing Fermi level. Why have we found this
particular type of structure? We know that the nanomechanical oscillator
potential seen by the electrons on the dot is independent of time when the
oscillator is in any of its pure eigenstates. Otherwise, when the oscillator
is not in a pure state, the potential seen by the electrons is time dependent.
In the former case, the electrons are scattered elastically by the time
independent potential and in the latter case the scattering process is
inelastic because the time dependent potential allow the transfer of energy
between the two. We observe that the constant average energy also has steps as
a function of the left Fermi level which become more pronounced with
increasing coupling strength. Hence, the oscillatory part of the behavior of
the mechanical oscillator is damped by coupling with the electrons on the dot
but the constant part is not. The damping mechanism in the transient dynamics
is due to transfer of energy from the nanomechanical oscillator to the
electrons on the dot while when the oscillator is in any of the pure
eigenstate then there is no mechanism for the transfer of energy between the
two. This same physical phenomenon also applies to the net current flowing
through the dot as well. This appear to be a specifically new quantum
phenomena in the study of nanomechanical systems.

\begin{figure}[h!tb]
\begin{center}
\includegraphics[width=0.5\textwidth]{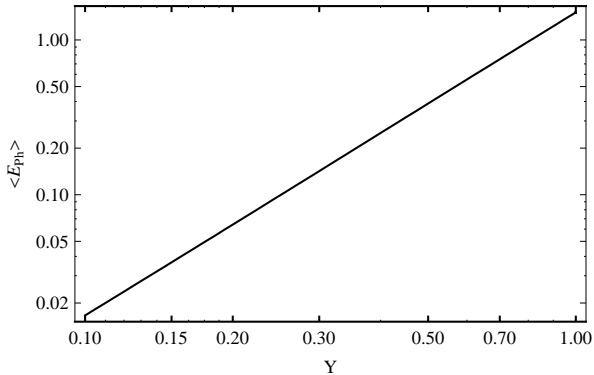}
\end{center}
\caption{\label{fig.7} Average energy transferred to the oscillator as a function of
$\frac{\hbar^{\prime}}{\hbar}$ and for fixed values of $\epsilon
_{0}=0.5,t=1000,\epsilon_{FR}=0,\epsilon_{FL}=1,\Gamma=1$ and $\eta=0.02$.
Units: all the parameters have same dimension as of $\hbar\omega$.}
\end{figure}

Can we compare this quantum phenomena with the classical mechanical
oscillator? Yes, the nanomechanical oscillator has to enter the classical
regime in the limit of small $\hbar$. For this, we study the dynamics of the
quantum oscillator in the classical limit, in which $\hbar$ in the
mechanical oscillator part of the Hamiltonian given in Eq.~(\ref{1}) goes to zero,
where $\hbar\omega<\Gamma$ . To see this, we have plotted the average energy
as a function of $Y=\frac{\hbar^{\prime}}{\hbar}$ in the nanomechanical
part of the system in fig.~\ref{fig.7} for fixed values of tunneling $\Gamma=1$,
$\epsilon_{\mathrm{F}R}=0,\epsilon_{\mathrm{F}L}=1$ and coupling strength
$\eta=0.05$. We found that the average energy of the quantum nanomechanical
oscillator scales as $\hbar^{2}$. We set the average energy in the limit
$\hbar^{\prime}\rightarrow0$ to see what happen to the system for long time.
It implies that in this limit, the energy transferred to the nanomechanical
oscillator is zero for long time. Hence, we conclude that the long time
dynamics of the classical mechanical oscillator is always zero.

\begin{figure}[h!tb]
\begin{center}
\includegraphics[width=0.5\textwidth]{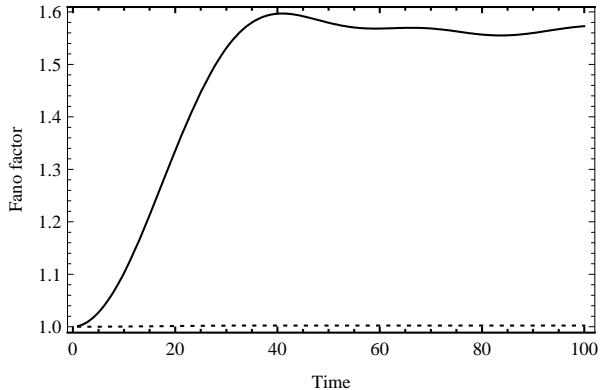}
\end{center}
\caption{\label{fig.8}Fano factor as a function of time for two different values of
coupling strength: $\eta=0.02$ (dotted line)$,$ and $0.08$ (solid line).
Parameters: $\epsilon_{0}=0.5,\epsilon_{FR}=0,\epsilon_{FL}=1,\hbar
\omega=0.1,\Gamma=0.01$. Units: all the parameters have same dimension as of
$\hbar\omega$.}
\end{figure}

Finally, in fig.~\ref{fig.8}, we have shown the Fano factor as a function of time for two different
values of $\eta=0.02,\eta=0.08$ and for fixed values of $\Gamma=0.01,\epsilon
_{\mathrm{F}R}=0,\epsilon_{\mathrm{F}L}=1$. In the limit of weak coupling, the
nanomechanical oscillator shows thermal like behavior and poissonian
statistics while in the limit of strong coupling its dynamics is non-thermal
which leads to super-poissonian statistics. In this figure, the short time
behavior is always thermal, but this is trivial as the nanomechanical
oscillator is initially in its ground state.

In conclusion, we have found mixed and pure states in our results which
confirm the quantum dynamics of our model with the following justifications:
in a classical mechanical oscillator model\cite{15,16,17,47} all states give
rise to a time dependent potential. Hence, all states of the classical
mechanical oscillator are damped. Thus, we confirm the new quantum dynamics of
the nanomechanical oscillator that will be helpful for further experiments
beyond the classical limit to develop better understanding of NEMS devices.

\section{Summary}
In this work, we analyzed the time-dependent quantum transport of a resonant
tunnel junction coupled to a nanomechanical oscillator by using the
nonequilibrium Green's function approach without treating the electron phonon
coupling as a perturbation. We have derived an expression for the full density
matrix or the dot population and discuss it in detail for different values of
the coupling strength and the tunneling rate. We derive an expression for the
current to see the effects of the coupling of the electrons to the oscillator
on the dot and the tunneling rate of electrons to resolve the dynamics of the
nanomechanical oscillator. This confirms that electrons resolve the dynamics
of nanomechanical oscillator in the regime $\tau_{e}>\tau_{\text{Osc}}$ while they do
not in the opposite case $\tau_{e}>\tau_{\text{Osc}}.$ Furthermore, we discuss the
average energy transferred to oscillator as a function of time. We also
discuss the Fano factor as a function of time, which shows thermal behavior
and poissonian to non-thermal and super-poissonian behavior. We have found new
dynamics of the nanomechanical oscillator: pure and mixed states, which are
never present in a classical oscillator. These results suggest further
experiments for NEMS to go beyond the classical dynamics.

\appendix
\section{\label{App.A}}
The particle current $I_{\alpha}$ into the interacting region from the lead is
related to the expectation value of the time derivative of the number operator
$N_{\alpha}=\sum_{\alpha j}c_{\alpha j}^{\dagger}c_{\alpha j}$, 
as\cite{25,35,36,37}%
\begin{eqnarray}
I_{\alpha}(t) &=&-e\langle\frac{d}{dt}x\rangle
=\frac{-\mathrm{i}e}{\hbar}\langle[H,x]\rangle \label{23}\\
I_{\alpha}(t) &=& \frac{e}{\hbar}\{\G_{0,\alpha}^{<}(t,t)V_{\alpha,0}%
(t)-V_{0,\alpha}^{\ast}(t)\G_{\alpha,0}^{<}(t,t)\}, \label{24}%
\end{eqnarray}
where we have the following relations%
\begin{eqnarray}
\G_{o,\alpha}^{<}(t,t) &=&%
%TCIMACRO{\dint }%
%BeginExpansion
{\displaystyle\int}
%EndExpansion
\mathrm{d}t^{\prime}\left\{%
\G_{0,0}^{r}(t,t^{\prime})V_{0,\alpha}(t^{\prime})\g_{\alpha,\alpha}^{<}(t^{\prime},t)
+\G_{0,0}^{<}(t,t^{\prime})V_{0,\alpha}(t^{\prime
})\g_{\alpha,\alpha}^{a}(t^{\prime},t)\right\} \label{25}\\
\G_{\alpha,0}^{<}(t,t) &=&%
%TCIMACRO{\dint }%
%BeginExpansion
{\displaystyle\int}
%EndExpansion
\mathrm{d}t^{\prime}\left\{%
\g_{\alpha,\alpha}^{r}(t,t^{\prime})V_{\alpha,0}(t^{\prime
})\G_{0,0}^{<}(t^{\prime},t)
+\g_{\alpha,\alpha}^{<}(t,t^{\prime})V_{\alpha
,0}(t^{\prime})\G_{0,0}^{a}(t^{\prime},t)\right\}, \label{26}%
\end{eqnarray}
where $\g_{\alpha,\alpha}^{r,(a),(<)}(t,t^{\prime})$ refers to the unperturbed
states of the leads and given as%
\[
\g_{\alpha,\alpha}^{r}(t,t^{\prime})
= \frac{1}{N}\sum_{j}\g_{\alpha,j}^{r}(t,t^{\prime})
= -\mathrm{i}n_{\alpha}\theta(t-t^{\prime})\overset{+\infty}%
{\underset{-\infty}{%
%TCIMACRO{\dint }%
%BeginExpansion
{\displaystyle\int}
%EndExpansion
}}\mathrm{d}\varepsilon_{\alpha}\exp[-\mathrm{i}\varepsilon_{\alpha}(t-t^{\prime})],
\]
with the fact that $\underset{j}{%
%TCIMACRO{\dsum }%
%BeginExpansion
{\displaystyle\sum}
%EndExpansion
}\mapsto\overset{+\infty}{\underset{-\infty}{%
%TCIMACRO{\dint }%
%BeginExpansion
{\displaystyle\int}
%EndExpansion
}}Nn_{\alpha}\mathrm{d}\varepsilon_{\alpha}$\ with $n_{\alpha}$ being the constant
number density of the leads and other uncoupled Green's function in the leads are%
\begin{eqnarray*}
\g_{\alpha,\alpha}^{a}(t,t^{\prime}) &=& \frac{1}{N}\underset{j}{%
%TCIMACRO{\dsum }%
%BeginExpansion
{\displaystyle\sum}
%EndExpansion
}\g_{\alpha,j}^{a}(t,t^{\prime})=+\mathrm{i}n_{\alpha}\theta(t^{\prime}-t)\overset
{+\infty}{\underset{-\infty}{%
%TCIMACRO{\dint }%
%BeginExpansion
{\displaystyle\int}
%EndExpansion
}}\mathrm{d}\varepsilon_{\alpha}\exp[-\mathrm{i}\varepsilon_{\alpha}(t-t^{\prime})] ,\\
\g_{\alpha,\alpha}^{<}(t,t^{\prime}) &=& \frac{1}{N}\underset{j}{%
%TCIMACRO{\dsum }%
%BeginExpansion
{\displaystyle\sum}
%EndExpansion
}\operatorname{f}_{\alpha}(\varepsilon_{\alpha})g_{\alpha,j}^{<}(t,t^{\prime})=\overset
{+\infty}{\underset{-\infty}{%
%TCIMACRO{\dint }%
%BeginExpansion
{\displaystyle\int}
%EndExpansion
}}\mathrm{d}\varepsilon_{\alpha}\operatorname{f}_{\alpha}(\varepsilon_{\alpha})\mathrm{i}n_{\alpha}%
\exp[-\mathrm{i}\varepsilon_{\alpha}(t-t^{\prime})],
\end{eqnarray*}
Now using equations~(\ref{25} \& \ref{26}) in the equation~(\ref{24}) of current through lead $\alpha$\ as%
\begin{eqnarray}
I_{\alpha}(t)  &=& \frac{e}{\hbar}%
%TCIMACRO{\dint }%
%BeginExpansion
{\displaystyle\int}
%EndExpansion
\mathrm{d}t^{\prime}\Tr\left\{\left(\G_{0,0}^{r}(t,t^{\prime})V_{0,\alpha}(t^{\prime})\g_{\alpha,\alpha}^{<}(t^{\prime},t)
+\G_{0,0}^{<}(t,t^{\prime})V_{0,\alpha}(t^{\prime})\g_{\alpha,\alpha}^{a}(t^{\prime},t)\right)V_{\alpha,0}(t)\right.\nonumber\\
&& -\left.V_{0,\alpha}^{\ast}(t)\left(\g_{\alpha,\alpha}^{r}(t,t^{\prime})V_{\alpha,0}(t^{\prime})\G_{0,0}^{<}(t^{\prime},t)
+\g_{\alpha,\alpha}^{<}(t,t^{\prime
})V_{\alpha,0}(t^{\prime})\G_{0,0}^{a}(t^{\prime},t)\right)\right\}, \label{27}%
\end{eqnarray}
Using the fact that $\Sigma_{0,0,\alpha}^{r,(a),(<)}(t^{\prime},t)=V_{0,\alpha
}^{\ast}(t^{\prime})\g_{\alpha,\alpha}^{r,(a),(<)}(t^{\prime},t)V_{\alpha
,0}(t)$, we can simplify the above equation as%
\begin{eqnarray}
I_{\alpha}(t)  &=& \frac{e}{\hbar}%
%TCIMACRO{\dint }%
%BeginExpansion
{\displaystyle\int}
%EndExpansion
\mathrm{d}t^{\prime}\Tr\left\{%
\G_{0,0}^{r}(t,t^{\prime})\Sigma_{0,0,\alpha}^{<}(t^{\prime},t)
+\G_{0,0}^{<}(t,t^{\prime})\Sigma_{0,0,\alpha}^{a}(t^{\prime},t)\right.\nonumber\\
&&  -\left.\Sigma_{0,0,\alpha}^{r}(t,t^{\prime})G_{0,0}^{<}(t^{\prime},t)
-\Sigma_{0,0,\alpha}^{<}(t,t^{\prime})G_{0,0}^{a}(t^{\prime},t)]\right\}, \label{28}%
\end{eqnarray}
where $\Sigma_{0,0,\alpha}^{r,(a),(<)}(t,t^{\prime})$ are non-zero only when
both the times ($t,t^{\prime}$) are positive $t,t^{\prime}>0$. Although
$\g^{r,(a)}(t,t^{\prime})$ is non-zero for $t<0$, it is never required due to
the way it combines with $\Sigma_{0,0,\alpha}^{r,(a),(<)}(t,t^{\prime})$. Here
we note that we require $\g^{r,(a)}(t,t^{\prime})$ from Eq.~(\ref{14} \& \ref{15}) for
positive times only ($t>0)$. The first integral on right hand side of Eq.~(\ref{28})
may be solved by using Eq.~(\ref{13}, \ref{14} \& \ref{17}) as%
\begin{eqnarray}
\hbox to 1.5cm{\rlap{$\displaystyle\Tr%
%TCIMACRO{\dint _{0}^{t}}%
%BeginExpansion
{\displaystyle\int_{0}^{t}}
%EndExpansion
\mathrm{d}t^{\prime}
\G_{0,0}^{r}(t,t^{\prime})\Sigma_{0,0,\alpha}^{<}(t^{\prime},t)
$}\hfil}\nonumber\\  
&=& \frac{-\Gamma}{2\pi}\sum_{m}%
%TCIMACRO{\dint \limits_{-\infty}^{\epsilon_{\mathrm{F}\alpha}}}%
%BeginExpansion
{\displaystyle\int\limits_{-\infty}^{\epsilon_{\mathrm{F}\alpha}}}
%EndExpansion
\mathrm{d}\varepsilon_{\alpha}%
%TCIMACRO{\dint _{0}^{t}}%
%BeginExpansion
{\displaystyle\int_{0}^{t}}
%EndExpansion
\mathrm{d}t^{\prime}\Phi_{0,m}\Phi_{0,m}^{\ast}
\exp[-\mathrm{i}(\varepsilon_{m}-\mathrm{i}\Gamma)(t-t^{\prime})]%
\exp[-\mathrm{i}\varepsilon_{\alpha}(t^{\prime}-t)]\nonumber\\
&=& \frac{\mathrm{i}\Gamma}{2\pi}\sum_{m}\Phi_{0,m}\Phi_{0,m}^{\ast}%
%TCIMACRO{\dint \limits_{-\infty}^{\epsilon_{\mathrm{F}\alpha}}}%
%BeginExpansion
{\displaystyle\int\limits_{-\infty}^{\epsilon_{\mathrm{F}\alpha}}}
%EndExpansion
\mathrm{d}\varepsilon_{\alpha}\left\{\frac{1-
\exp[\mathrm{i}(\varepsilon_{\alpha}-\varepsilon_{m}+\mathrm{i}\Gamma)t]}%
{\varepsilon_{\alpha}-\varepsilon_{m}+\mathrm{i}\Gamma}\right\}\nonumber\\
&=& \frac{\mathrm{i}\Gamma}{2\pi}\sum_{m}\Phi_{0,m}\Phi_{0,m}^{\ast}\left\{\ln(\epsilon_{\mathrm{F}\alpha}-\varepsilon_{m}+\mathrm{i}\Gamma) -\Ei[\mathrm{i}(\epsilon_{\mathrm{F}%
\alpha}-\varepsilon_{m}+\mathrm{i}\Gamma)t]\right\}, \label{29}%
\end{eqnarray}
where the final result is obtained using standard integrals\cite{45}.
We note once again that special care is required in evaluating the $\ln(x)$ and $\Ei(x)$ to choose
the correct Riemann sheets in order to make sure that these functions are
consistent with the initial conditions and are continuous functions of time
and chemical potential. This statement will also apply to all further discussions.

The second \& third integral on right hand side of Eq.~(\ref{28}) are written as%
\[
\Tr%
%TCIMACRO{\dint _{0}^{t}}%
%BeginExpansion
{\displaystyle\int_{0}^{t}}
%EndExpansion
\mathrm{d}t^{\prime}\left\{%
\G_{0,0}^{<}(t,t^{\prime})\Sigma_{0,0,\alpha}^{a}(t^{\prime},t)
-\Sigma_{0,0,\alpha}^{r}(t,t^{\prime})\G_{0,0}^{<}(t^{\prime},t)\right\}
=\mathrm{i}\Gamma\Tr\G_{0,0}^{<}(t,t).
\]
This integral can be solved in the same way as for the dot population. The final
result is written as\cite{45}%
\begin{eqnarray}
\mathrm{i}\Gamma\Tr\G_{0,0}^{<}(t,t)  
&=& \frac{\Gamma}{2\pi}\sum_{\alpha,m}\Phi_{0,m}\Phi_{0,m}^{\ast}\biggl\{
-\left(1+\exp[-2\Gamma t]\right)%
\left(\tan^{-1}\left[\frac{\epsilon_{\mathrm{F}\alpha}%
-\varepsilon_{m}}{\Gamma}\right]+\frac{\pi}{2}\right)\nonumber\\
&&  +\half\mathrm{i}\exp[-2\Gamma t]\left(%
-\Ei[+\mathrm{i}(\epsilon_{\mathrm{F}\alpha}-\varepsilon_{m}-\mathrm{i}\Gamma)t]
+\Ei[-\mathrm{i}(\epsilon_{\mathrm{F}\alpha}-\varepsilon_{m}+\mathrm{i}\Gamma)t]\right)\nonumber\\
&&  +\half\mathrm{i}\left(%
 \Ei[+\mathrm{i}(\epsilon_{\mathrm{F}\alpha}-\varepsilon_{m}+\mathrm{i}\Gamma)t]
-\Ei[-\mathrm{i}(\epsilon_{\mathrm{F}\alpha}-\varepsilon_{m}-\mathrm{i}\Gamma)t]%
\right)\biggr\}, \label{30}%
\end{eqnarray}
and the fourth integral on right hand side of equation~(\ref{28}) can be solved by using Eq.~(\ref{13}, \ref{15}, \& \ref{17}) as%
\begin{eqnarray}
\hbox to 1.5cm{\rlap{$\displaystyle-\Tr%
%TCIMACRO{\dint _{0}^{t}}%
%BeginExpansion
{\displaystyle\int_{0}^{t}}
%EndExpansion
\mathrm{d}t^{\prime}\Sigma_{0,0,\alpha}^{<}(t,t^{\prime})\G_{0,0}^{a}(t^{\prime},t)$}\hfil}\nonumber\\
&=& \frac{\Gamma}{2\pi}\sum_{m}%
%TCIMACRO{\dint \limits_{-\infty}^{\epsilon_{FL}}}%
%BeginExpansion
{\displaystyle\int\limits_{-\infty}^{\epsilon_{\mathrm{F}\alpha}}}
%EndExpansion
d\varepsilon_{\alpha}%
%TCIMACRO{\dint _{0}^{t}}%
%BeginExpansion
{\displaystyle\int_{0}^{t}}
%EndExpansion
\mathrm{d}t^{\prime}\Phi_{0,m}\Phi_{0,m}^{\ast}
\exp[-\mathrm{i}(\varepsilon_{m}+\mathrm{i}\Gamma)(t^{\prime}-t)]\exp
[-\mathrm{i}\varepsilon_{\alpha}(t-t^{\prime})]\nonumber\\
&=&\frac{-\mathrm{i}\Gamma}{2\pi}\sum_{m}\Phi_{0,m}\Phi_{0,m}^{\ast}%
%TCIMACRO{\dint \limits_{-\infty}^{\epsilon_{\mathrm{F}\alpha}}}%
%BeginExpansion
{\displaystyle\int\limits_{-\infty}^{\epsilon_{\mathrm{F}\alpha}}}
%EndExpansion
\mathrm{d}\varepsilon_{\alpha}\left\{\frac{1-\exp[-\mathrm{i}(\varepsilon_{\alpha}-\varepsilon
_{m}-\mathrm{i}\Gamma)t}{\varepsilon_{\alpha}-\varepsilon_{m}-\mathrm{i}\Gamma}\right\}\nonumber\\
&=& \frac{-\mathrm{i}\Gamma}{2\pi}\sum_{m}\Phi_{0,m}\Phi_{0,m}^{\ast}\left\{
\ln(\epsilon_{\mathrm{F}\alpha}-\varepsilon_{m}-\mathrm{i}\Gamma)
-\Ei[-\mathrm{i}(\epsilon_{\mathrm{F}\alpha}-\varepsilon_{m}-\mathrm{i}\Gamma)t]
\right\}  \label{31}%
\end{eqnarray}
Using equations~(\ref{29}, \ref{30} \& \ref{31}) in Eq.~(\ref{28}), the final expression for the current is written as%
\begin{equation}
I_{\alpha}(t)=\frac{e\Gamma}{2\pi\hbar}\sum_{m}\Phi_{0,m}\Phi_{0,m}^{\ast
}\{I^{1\alpha}_m+I^{2L}_m+I^{2R}_m\}, \label{32}%
\end{equation}
where components of current are written as%
\begin{eqnarray*}
I^{1\alpha}_m &=& 2\left(\tan^{-1}\left[\frac{\epsilon_{\mathrm{F}\alpha}-\varepsilon_{m}%
}{\Gamma}\right]+\frac{\pi}{2}\right)\\
&&-\mathrm{i}\left\{%
 \Ei[+\mathrm{i}(\epsilon_{\mathrm{F}\alpha}-\varepsilon_{m}+\mathrm{i}\Gamma)t]
-\Ei[-\mathrm{i}(\epsilon_{\mathrm{F}\alpha}-\varepsilon_{m}-\mathrm{i}\Gamma)t]
\right\}],\\
I^{2\alpha}_m  &=& -\left(1+\exp[-2\Gamma t]\right)
\left\{\tan^{-1}\left[\frac{\epsilon_{\mathrm{F}\alpha}-\varepsilon_{m}}{\Gamma}\right]+\frac{\pi}{2}\right\}\\
&&  -\half\mathrm{i}\exp[-2\Gamma t]\left\{
 \Ei[+\mathrm{i}(\epsilon_{\mathrm{F}\alpha}-\varepsilon_{m}-\mathrm{i}\Gamma)t]
-\Ei[-\mathrm{i}(\epsilon_{\mathrm{F}\alpha}-\varepsilon_{m}+\mathrm{i}\Gamma)t]
\right\}\\
&&  +\half\mathrm{i}\left\{
 \Ei[+\mathrm{i}(\epsilon_{\mathrm{F}\alpha}-\varepsilon_{m}+\mathrm{i}\Gamma)t]
-\Ei[-\mathrm{i}(\epsilon_{\mathrm{F}\alpha}-\varepsilon_{m}-\mathrm{i}\Gamma)t]
\right\},
\end{eqnarray*}
where in calculating the left current we need $I^{1L}_m$ together with both
$I^{2L}_m$ and $I^{2R}_m$ whereas for the right current $I^{1L}_m$ is replaced by $I^{1R}_m$.

\begin{acknowledgement}
M.Tahir would like to acknowledge the support of the Pakistan Higher Education
Commission (HEC).
\end{acknowledgement}

\end{document}